\documentclass{emulateapj}

\slugcomment{{\sc Submitted to ApJ:} September 14, 2009}
\newcommand{\msol}{\hbox{$M_\odot$}}

\shorttitle{Correlation of SNR Masers and $\gamma$-ray Sources}
\shortauthors{Hewitt, Yusef-Zadeh and Wardle}

\begin{document}

\title{Correlation of Supernova Remnant Masers and Gamma-Ray Sources}
\author{John W. Hewitt, Farhad Yusef-Zadeh}
\affil{Department of Physics and Astronomy, Northwestern University, Evanston, IL 60208}
\and
\author{Mark Wardle}
\affil{Department of Physics and Engineering, Macquarie University, Sydney, NSW 2109, Australia}

\begin{abstract}
Supernova remnants interacting with molecular clouds are potentially exciting systems in which to detect evidence of cosmic ray acceleration. Prominent $\gamma$-ray emission is produced via the decay of neutral pions when cosmic rays encounter the nearby dense clouds. In many of the supernova remnants coincident with $\gamma$-ray sources, the presence of OH(1720 MHz) masers is used to identify interaction with dense gas and to provide a kinematic distance to the system. In this paper we use statistical tests to demonstrate that there is a correlation between these masers and a class of GeV- to TeV-energy $\gamma$-ray sources coincident with interacting remnants. For pion decay the $\gamma$-ray luminosity provides a direct estimate of the local cosmic ray density. We find the cosmic ray density is enhanced by one to two orders of magnitude over the local solar value, comparable to X-ray-induced ionization in these remnants. The inferred ionization rates are sufficient to explain non-equilibrium chemistry in the post-shock gas, where high columns of hydroxyl are observed.
\end{abstract}
\keywords{cosmic rays --- masers --- supernova remnants --- shock waves}

\section{Introduction}

It has long been speculated that the vast energy of supernova remnants (SNRs) is tapped to produce the bulk of Galactic cosmic rays. Those SNRs interacting with molecular clouds are excellent targets to detect evidence of cosmic ray acceleration. The nearby dense gas acts as a target for hadronic cosmic rays, producing $\gamma$-rays via the decay of neutral pions (e.g., Drury et al. 1994).

Interest in cosmic rays from interacting SNRs has been reinvigorated by recent detections of $\gamma$-ray emission at GeV- to multi-TeV-energies from supernova remnants. The HESS galactic plane survey yielded 14 new TeV detections, 7 of which are coincident with SNRs \citep{aharonian06}. Targeted observations are now capable of reaching sufficient sensitivities to resolve the morphology of $\gamma$-ray emission associated with the remnant \citep{hess_ctb37a,hess_w28}. In particular, a TeV-energy counterpart to IC 443 was observed to be spatially offset from the GeV-energy $\gamma$-ray source \citep{magic_ic443,veritas_ic443}. The spatial variations with energy may indicate the diffusion of cosmic rays accelerated by the supernova remnant out into the adjacent molecular cloud \citep{torres08}.

Interacting SNRs represent a promising class of $\gamma$-ray sources which are likely to be uncovered in increasing numbers by the current generation of $\gamma$-ray observatories. However, to date the identification of $\gamma$-ray counterparts to SNRs has been inhibited by the poor spatial resolution and sensitivity of $\gamma$-ray telescopes \citep{esposito96,torres03}. Often, the association of $\gamma$-ray sources with interacting SNRs rests on supplemental evidence from common tracers of interaction with dense gas: broad molecular lines, bright at infrared to millimeter wavelengths (e.g., Reach et al. 2005), and the hydroxyl(OH) maser at 1720 MHz (e.g., Frail et al. 1994).

OH masers, when detected only at 1720 MHz, are a particularly powerful diagnostic which is only associated with SNR shocks. While these "SNR masers" are only detected in $\sim$10\% of remnants, they unambiguously trace cool (25--150 K), dense ($\sim$10$^5$ cm$^{-3}$) gas in the wake of non-dissociative shocks and permit a direct measure of magnetic field strength \citep{lockett99}. Prominent interacting SNRs W28, W44 and IC 443 have been noted for both bright masers and coincident EGRET sources, leading to the suggestion that masers may trace cosmic-ray acceleration sites \citep{claussen97}.

In this letter we address the nature of interacting SNRs with $\gamma$-ray counterparts: First, we show that an excellent correlation exists between SNR masers and a subset of GeV- and TeV-energy sources toward SNRs. This strengthens the argument for a hadronic origin for the observed $\gamma$-rays. Second, if pion decay is indeed the dominant emission mechanism, the implied cosmic ray enhancement is sufficient to explain the increased ionization needed to produce high columns of OH in the post-shock gas \citep{wardle99}. This suggests there is a more intimate relationship between SNR masers and $\gamma$-ray sources than merely being complimentary tracers of dense gas interaction. 

\begin{deluxetable*}{rrlrrrrrrr}
\tablecaption{Known SNR Masers and Coincident $\gamma$-ray Sources}
\tablewidth{0pt}
\tabletypesize{\scriptsize}
\tablehead{
\colhead{$l$} & \colhead{$b$} & \colhead{SNR} & \colhead{Diameter} & \colhead{Distance} &
\colhead{L$_{GeV}$} & \colhead{$\alpha_{GeV}$} &
\colhead{L$_{TeV}$} & \colhead{$\alpha_{TeV}$} & Ref. \\
& & & \colhead{(\arcmin )} & \colhead{(kpc)} & \colhead{(ergs s$^{-1}$)} & & \colhead{(ergs s$^{-1}$)}
}
\startdata
\multicolumn{10}{c}{Group A}\\
\hline
6.4		&--0.1	&W28 		&42		&2.0			&4.8e35	&2.1		&1.5e33	&2.7 & 1,2\\
34.7		&--0.4	&W44 		&30		&2.5			&4.1e35	&1.9		&---		&--- & 1 \\
49.2    &--0.7	&W51 C 		&30		&6			&1.7e36	&--		&1.5e33	&--  & 3,4 \\ 
189.1	&+3.0	&IC 443 		&50		&1.5			&1.0e35	&2.0		&1.2e33	&3.1 & 1,5,6\\
\hline
\multicolumn{10}{c}{Group B}\\
\hline
0.0		&+0.0	&SgrA East	&2.5		&8.5			&1.2e37	&1.7		&1.3e36	&2.2 & \\
5.7		&--0.0	&    		&9		&3.2			&---		&---		&0.8e33	&2.3 & 2 \\
8.7		&--0.1	&W30 		&45		&3.9			&---		&---		&1.7e34	&2.7 & \\
337.8	&--0.1	&Kes 41		&5		&12.3		&4.2e36	&2.5		&---		&--- & \\
348.5	&+0.1	&CTB 37A 	&10		&11.3		&3.5e36	&2.3		&5.3e33	&2.3 & 7\\
359.1	&--0.5	&			&10		&5.0			&1.2e36	&2.2		&7.5e32	&2.7 & 8\\
\hline
\multicolumn{10}{c}{Group C}\\
\hline
  1.0	&--0.1	&Sgr D SNR&8		&8.5	&	\\
  1.4	&--0.1	&&10		&8.5	&	\\
  5.4	&--1.2	&Duck	&35		&5.2	&	\\
  9.7	&--0.0	&&11		&4.7	&	\\
 16.7	&+0.1	&&4		&2/14	&	\\
 21.8	&--0.6	&Kes 69	&20		&5.2	&	\\
 31.9	&--0.0	&3C 391	&8		&9		&	\\
 32.8	&--0.1	&Kes 78	&20		&5.5/8.5&	\\
337.0	&--0.1	&CTB 33	&3		&11		&	\\
346.6	&--0.2	&&8		&11		&	\\
348.5	&--0.0	&&10		&13.7	&	\\
349.7	&+0.2	&&2		&$>$11	&	\\
357.7	&+0.3	&Square	&24		&6.4	&	\\
357.7	&--0.1	&Tornado	&5		&$>$6		
\enddata
\tablecomments{
{\scriptsize EGRET luminosity from 0.04-6 GeV and HESS or VERITAS luminosity from 0.3-10 TeV, where $\alpha$ is the photon index.
\noindent REFERENCES. -- (1) \citet{esposito96}, (2) \citet{hess_w28}, (3) \citet{abdo09}, (4) \cite{feinstein09}, (5) \cite{veritas_ic443}, (6) \citet{magic_ic443}, (7) \citet{hess_ctb37a}, (8) \cite{aharonian06} \label{tbl:list}} }
\end{deluxetable*}

\section{$\gamma$-ray Emission from Interacting SNRs}
First, we assess $\gamma$-ray sources which may be directly associated with supernova remnants (hereafter, "$\gamma$-ray SNRs).  To identify potential $\gamma$-ray counterparts to SNRs we have drawn from three catalogs: EGRET sources \citep{torres03}, the $Fermi$ Bright Source List \citep{abdo09} and the TeVCat\footnote{TeVCat is an online catalog of very-high-energy ($>$50 GeV) $\gamma$-ray sources. It can be accessed at http://tevcat.uchicago.edu .}. It is not trivial to determine the origins of each $\gamma$-ray source. Potential associations can be confused by the presence of pulsars or pulsar wind nebulae, which are also capable of producing high-energy $\gamma$-ray emission. Some sources reported by \citet{torres03} as potential EGRET counterparts to SNRs have since been associated with other astrophysical sources. The source 3EG J1410-6147, coincident with SNR G312.4-0.4, has recently been associated with the young pulsar J1410-6132 with variability detected by AGILE \citep{obrien08}. The source 3EG J1824-1514 has been confirmed as the microquasar LS 5039 \citep{ls5039}, ruling out an association with SNR G16.8-1.1 \citep{torres03}. Sources 3EG J2016+3657 and 3EG J2020+4017, coincident with SNRs G74.9+1.2 and G78.2+2.1 respectively, are identified as extragalactic blazars \citep{iyudin07}. Finally, the EGRET counterparts for SNRs G180.0-1.7, G355.6+0.0, G39.2-0.3 and G74.8+1.2 \citep{torres03} do not appear in the Fermi bright source catalog \citep{abdo09}, and we therefore do not include them as detections in our analysis. 

We include all remnants for which an association with the coincident $\gamma$-ray source has not been ruled out. There are 26 identified SNRs with $\gamma$-ray coincidences: 7 are young remnants ($\la$1 kyr), 12 are SNRs with evidence of interaction with dense clouds, and 7 are unclassified remnants. Of the 12 identified SNRs which are interacting with dense gas, all but two (MSH 11-61A and W49B) have detected SNR masers. The presence of masers gives several advantages: (1) masers signpost interaction with dense (10$^5$ cm$^{-3}$) clouds, which will enhance the pion decay signature, (2) the velocity of the maser gives a kinematic distance, and (3) an established velocity allows the adjacent cloud to be isolated from confusing Galactic emission along the line of sight. 

Table \ref{tbl:list} lists all SNRs with masers and the properties of potentially associated $\gamma$-ray emission. In the first five columns the Galactic coordinates, name, diameter and kinematic distance are listed for each remnant. The $\gamma$-ray luminosity for all derived from the reported $\sim$100 MeV--10 GeV and $\gtrsim$1 TeV fluxes in columns 6 and 8. Spectral indices for are given in columns 7 and 9. References for detections are given in the last column. 

The ten SNRs with coincident $\gamma$-ray sources are divided based on the certainty of their association: Group A includes four SNRs for which $\gamma$-rays are established as related to the SNR; Group B includes six SNRs with coincident $\gamma$-ray sources which have not been attributed to other astrophysical phenomenon (pulsars, blazars, etc.) but for which an association with the SNR is less than certain; Group C lists the sixteen SNRs with masers which do not yet have detected $\gamma$-ray sources. Given that both masers and $\gamma$-rays are detected for only 10$\%$ of SNRs, the large number of coincident detections makes an association between SNR masers and $\gamma$-ray emission as tracers of interaction quite plausible. 

\begin{deluxetable*}{rrrrrrrrrrrrr}
\label{tbl:contingency}
\tablecaption{Contingency Table: Presence of SNR Masers versus SNRs with $\gamma$-ray Sources \label{tbl:contingency}}
\tablewidth{0pt}
\tabletypesize{\scriptsize}
\tablehead{ 
  \colhead{}    &  
  \multicolumn{3}{c}{$\gamma$-ray SNRs} &&
  \multicolumn{3}{c}{No $\gamma$-ray Counterpart} \\ 
  \cline{2-4} \cline{6-8} \\ 
  \colhead{} & \colhead{} & \colhead{Expected} & \colhead{} && & \colhead{Expected} & \colhead{} \\
  \colhead{OH (1720 MHz)} & & \colhead{from} & \colhead{} && \colhead{} & \colhead{from} & \colhead{} & \colhead{} \\
  \colhead{Masers} & \colhead{Number} & \colhead{Counts} & \colhead{$\chi^2$} && \colhead{Number} & \colhead{Counts} & \colhead{$\chi^2$} & \colhead{Total}
}
\startdata
Masers present ....... & 10 &  2.4 & 24.7 &&  14 & 21.6 & 2.7 & 24\\
Masers absent ........ & 14 & 19.5 &  1.6 && 185 & 179.0& 0.2 & 199\\
No observations....... &  2 &  4.1 &  1.1 &&  41 & 37.9 & 0.1& 43 \\
Total ................ & 26 & & && 240 & & & 266 &
\enddata
\tablecomments{We do not include those 43 SNRs which have not been observed for SNR Masers.}
\end{deluxetable*}

\subsection{Contingency Table Analysis}

To explore the correlation between SNR masers and $\gamma$-ray sources, we use a contingency table analysis to test the null hypothesis that there is no association between the two groups. Here we include all remnants which have been searched for SNR masers \citep{frail96,green97,koralesky98,fyz99,hewitt09} and those $\gamma$-ray sources which have not been clearly identified as having a non-SNR origin. 

Table \ref{tbl:contingency} is the resulting contingency table. Three rows are given for SNRs with masers present, absent or not surveyed. Two columns divide SNRs with and without coincident $\gamma$-ray detections. Each of the two columns lists the "Number" of SNRs which meet both classifications, the number "Expected from Counts" assuming the maser and $\gamma$-ray properties of remnants are completely independent, and "Contribution to $\chi^2$" which gives the $\chi^2$, (observed -- expected)$^2$/(expected). For this test the total $\chi^2$ is  31 with two degrees of freedom. This gives a probability of $<$0.0001$\%$ that SNR masers and $\gamma$-ray counterparts to SNRs are not correlated.

We note that two of the expected cell frequencies are less than five. To be statistically rigorous we also apply the Fisher Exact Probability Test in addition to the $\chi^2$ Test. The Fisher test considers all possible outcomes of the contingency table in order to determine the probability of finding a correlation at least as strong as is observed. Here again we find a clear rejection of independence between $\gamma$-ray and maser detections in SNRs, with a probability of 0.002$\%$. The two classes are clearly associated. 

A similar contingency table analysis was used to quantify an association between SNR masers and mixed-morphology remnants, a class of SNRs with shell-type radio morphologies with prominent thermal X-ray emission from their interiors \citep{fyz03mmsnr}. Both classes of remnants are thought to result from interaction with dense gas and are strongly correlated. Of the identified $\gamma$-ray SNRs, 9 are classified as both mixed-morphology and maser-emitting, 2 are mixed-morphology without detected masers (SNRs W49B and MSH 11-61A), and only SNR G5.7-0.0 has detected masers but no X-ray detection. Therefore, either classification produces a nearly identically good correlation with $\gamma$-ray SNRs. Both X-rays and cosmic rays are capable of providing the ionization needed to produce SNR masers. Detailed discussion is given in Section \ref{sec:ionization}.

An additional complication arises in that many SNRs have associated pulsars with their own wind nebulae that are also capable of accelerating particles to TeV energies. To account for this we separate the $\gamma$-ray SNRs into two categories based on whether a PWN is or is not also detected. This reduces the number of sources that fall under each classification, lowering the probability of independence from the Fisher test to 1.2$\%$, still significant enough to reject the null hypothesis. Furthermore, there is evidence that at least four SNRs (see Table \ref{tbl:list}, "Group A") are associated with $\gamma$-ray emission from the SNR-cloud interaction, and not the PWN. Improved spatial resolution is needed to discriminate between $\gamma$-ray emission from the PWN and SNR, but we note that there are no markedly different characteristics between the two groups of SNRs with and without PWN. 

\subsection{Properties of $\gamma$-ray SNRs with Masers}

For SNRs interacting with dense gas, $\gamma$-ray counterparts are likely to be dominated by emission from neutral pion decay originating from interactions between hadronic cosmic rays and gas nuclei. In contrast to young SNRs, older interacting SNRs have little if any detected X-ray synchrotron emission. An inverse-Compton scenario requires small magnetic field strengths $<$1 $\mu$G \citep{yamazaki06} whereas Zeeman splitting of SNR masers measures field strengths of order $\sim$1 mG \citep{fyz96,claussen97,brogan00}.
Electronic bremsstrahlung emission has also been proposed, but the expectation would be for $\gamma$-ray emission to trace the radio shell morphology \citep{bykov00}.  TeV emission from W28 and IC 443 shows no correlation with the radio shell, and an excellent correlation with dense gas \citep{hess_w28,veritas_ic443}. Forthcoming Fermi observations will have sufficient resolution to resolve this question at GeV energies.

\begin{figure}
\centering
\includegraphics[height=1.5in]{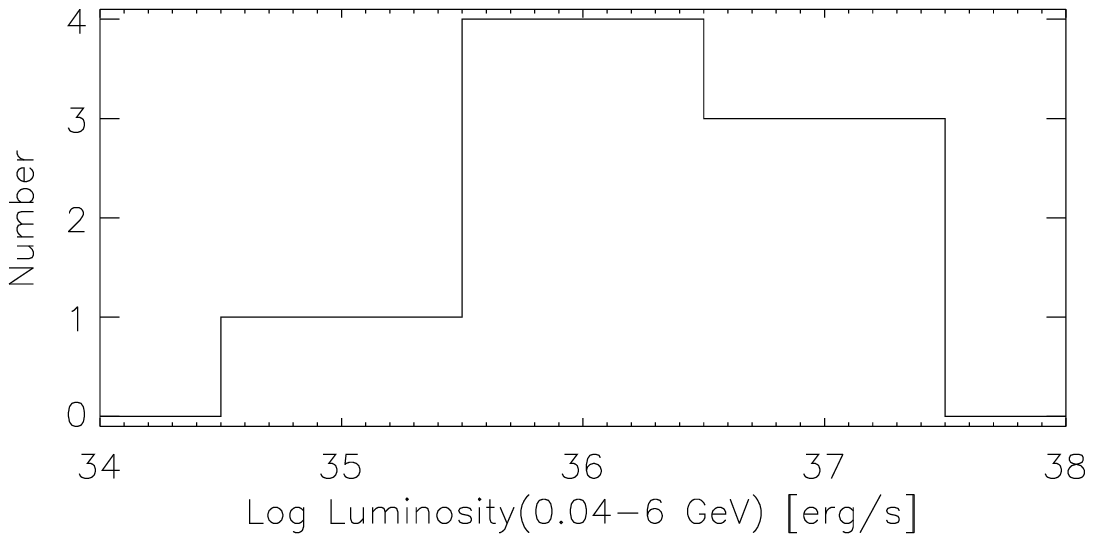}\\
\includegraphics[height=1.5in]{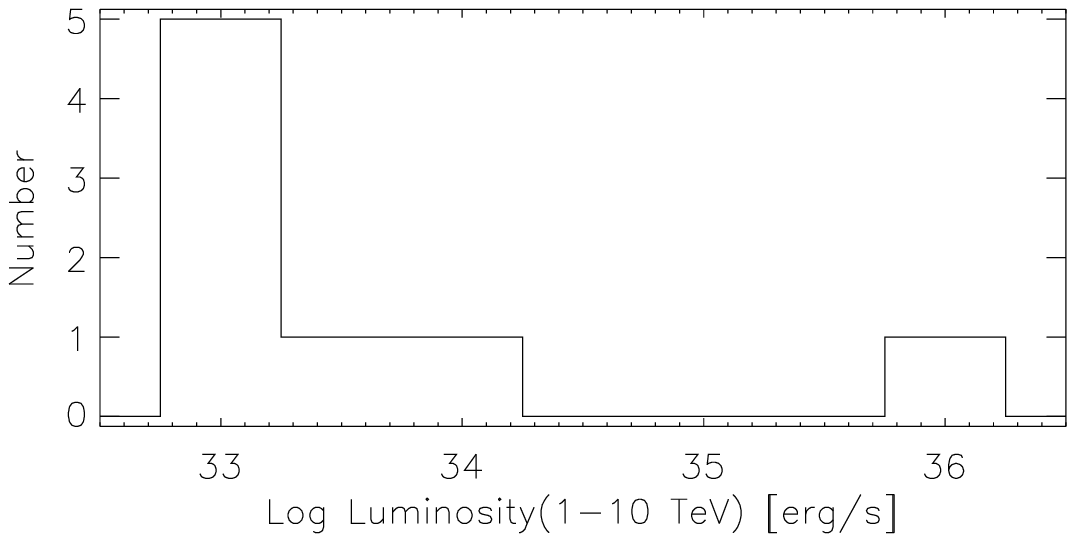}
\caption{Histograms of GeV (top) and TeV (bottom) luminosities for SNRs with masers. Bin sizes of 1 and 0.5 are used. Characteristic luminosities of 1.7$\times$10$^{36}$ ergs s$^{-1}$ and 1.5$\times$10$^{33}$ ergs s$^{-1}$ are found, respectively. 
\label{fig:histo}}
\end{figure}

The presence of a large reservoir of dense gas around the SNR, the expectation that cosmic rays are accelerated by SNRs, and the relatively older ages of this subset of interacting SNRs supports a pion decay origin for $\gamma$-ray counterparts. The morphology of $\gamma$-ray emission is seen to be well matched to that of the molecular cloud. In Figure \ref{fig:histo}, histograms of $\gamma$-ray luminosity show medians of 1.7$\times$10$^{36}$ ergs s$^{-1}$ and 1.5$\times$10$^{33}$ ergs s$^{-1}$ for GeV and TeV detections, respectively. This is consistent  with luminosity estimates if a significant fraction of the SN energy (10-20\%) is diverted to cosmic-rays which are incident on the adjacent molecular cloud \citep{drury94}. The orders of magnitude differences in the luminosity reflect the expected energy spectrum of accelerated particles. 

On average a hardening of the photon spectrum from GeV to TeV energies is also observed. The $\gamma$-ray spectral index, given in columns 7 and 9 of Table \ref{tbl:list}, steepen at higher energies from a mean of $\alpha_{GeV} \sim$ 2.1 to $\alpha_{TeV} \sim$ 2.6. It has been suggested for nearby SNR IC 443 that this spectral steepening results either from the presence of both pion decay and a strong bremsstrahlung component \citep{bykov00}, or the diffusion of cosmic rays at different energies through the dense cloud \citep{torres08}. In addition to IC 443, the SNRs W28, Sgr A East, and G359.1-0.5 show this characteristic spectral hardening at higher energies, but CTB 37A does not. Future detailed spectral modeling of these sources can determine whether a spectral signature can be used to discriminate between different  $\gamma$-ray emission mechanisms.

\section{Enhanced Ionizations via Cosmic Rays\label{sec:ionization}}
\begin{deluxetable*}{rrrrrrrrrrr}
\tablecaption{Comparison of Cosmic Ray and X-ray Ionizations
\label{tbl:crionization}}
\tablewidth{0pt}
\tabletypesize{\scriptsize}
\tablehead{
\colhead{$SNR$} & \colhead{Distance} & \colhead{M$_{cloud}$} & 
\colhead{F$_{\gamma}$($>$100 MeV)} & \colhead{$\zeta_{CR}$ } &
 \colhead{$\zeta_{Xray}$ } & \colhead{Ref.} \\
& \colhead{(kpc)} & \colhead{(10$^5$ $\msol$)} & \colhead{(10$^{-8}$ cm$^{-2}$ s$^{-1}$)} & \colhead{(10$^{-16}$ s$^{-1}$)} & \colhead{(10$^{-16}$ s$^{-1}$)} 
}
\startdata
W28 				&2.0			&0.5 &74.2 & 3.4 & 2.3 & 1\\
W44 				&2.5			&3.0 &88.9 & 1.1 & 4.5 & 2\\
W51 C 			&6.0			&1.9 &40.9 & 4.4 & 8.8 & 3\\
IC 443 			&1.5			&0.1 &51.4 & 5.5 & 3.6& 4\\
\enddata
\tablecomments{References: 
(1) \citet{hess_w28},
(2) \citet{seta98},
(3) \citet{feinstein09},
(4) \citet{dickman92}. Fluxes are taken from the Fermi Bright Source List \citep{abdo09}.}
\end{deluxetable*}

The strong correlation observed between $\gamma$-ray-bright and maser SNRs may result from more than just both tracing interaction with high gas densities. The large post-shock OH abundances required for masing to occur are not produced in chemical models of slow, dense shocks; instead all gas-phase oxygen is rapidly converted to water before the gas cools to 50 K \citep{kaufman96}. 

The large observed OH column densities are thought to be produced by dissociation of the abundant post-shock water \citep{wardle99}. In this model energetic electrons produced by ionizations in the molecular gas will excite H$_2$, which will subsequently generate a weak flux of far-ultraviolet photons which photo-dissociate water molecules producing OH \citep{prasad83}. The high flux of X-rays from the hot interior characteristic of mixed morphology SNRs has been shown to be a viable ionization mechanism \citep{fyz03mmsnr}. Cosmic rays and X-rays are notionally lumped together in this model.

The $\gamma$-ray luminosity can be used to discriminate the relative importance of ionizations by cosmic rays. The $\gamma$-ray spectrum resulting from pion decay of hadronic cosmic ray interactions in SNRs has been described by many authors \citep{drury94,baring99,bykov00}. 
Here we follow \citet{aharonian91}, formulating the $\gamma$-ray flux as a function of the  cosmic ray enhancement in the cloud adjacent to SNR, neglecting any possible gradients through the cloud as a simplification. The $\gamma$-ray flux above energy E$_{\gamma}$ is given by the equation:
\begin{equation}
{\rm 
F( \geq E_\gamma) \simeq 3\times10^{-13} \left(\frac{E_\gamma}{TeV}\right)^{-1.6} \omega_{CR} \left(\frac{M_{5}}{d_{kpc}^2} \right) 
 ph\ cm^{-2}\ s^{-1}
},
\label{eqn:1}
\end{equation}
where $d_{kpc}$ is the distance to the SNR in kpc, E$_\gamma$ is the threshold above which the $\gamma$-ray flux is totaled, $\omega_{CR}$ is the ratio of cosmic ray density at the SNR to the local solar value assuming the same $\gamma$-ray emissivity, and M$_5$ is the total mass of the SNR shell and adjacent molecular cloud in units of 10$^5$ \msol . We note that \citet{drury94} established that spectral variations between 2.0 and 2.7 changes this estimate by less than 20\% . 

Assuming the observed $\gamma$-ray flux from the SNR is due to hadronic interactions with the surrounding dense gas, it is straightforward to roughly estimate the local energy density of cosmic rays, provided the distance and molecular environment are known.  For EGRET and Fermi detections, E$_{\gamma}$ $>$ 100 MeV, equation (\ref{eqn:1}) can be rewritten to express the local cosmic ray enhancement $\omega_{CR}$:
\begin{equation}
\omega_{CR} \simeq\ \frac{F_{\gamma}(> 100 \ MeV)}{7 \times\ 10^{-7} {\rm\ ph \ cm^{-2}\ s^{-1}}} \ \left(\frac{d_{kpc}^2}{M_5}\right) \ .
\end{equation}
The cosmic ray ionization scales as the cosmic ray density, so we obtain the ionization rate due to cosmic rays $\zeta_{CR} \simeq\ \omega_{CR}\ \zeta_{\odot}$, where $\zeta_{\odot}$ is the local cosmic ray ionization rate of $\sim$4$\times$10$^{-17}$ s$^{-1}$ \citep{webber98}. 

In Table \ref{tbl:crionization} we list the observed properties of SNRs from which the local enhancement of cosmic rays is estimated. We find cosmic ray densities are typically increased by 30--150 times that observed for quiescent molecular clouds. This provides an ionization rate sufficient to produce OH abundances of $\sim$10$^{-7}$--10$^{-6}$ in the post-shock gas.
For comparison we also list the ionization rate from interior X-ray emission derived by \citet{fyz03mmsnr}. The two sources of ionization are generally found to be comparable. Either could be the dominant ionization mechanism depending on the location of the gas with respect to the interior X-ray emitting plasma and cosmic ray acceleration sites, and may vary on a source-by-source basis.

We caution that our estimates of cosmic ray ionization rate are somewhat crude. Even if the dominant emission mechanism is hadronic, there may still be a significant bremsstrahlung component. Detailed spectral modeling with data from the $Fermi$ $\gamma$-ray Space Telescope will permit a much better estimate of the local cosmic ray density. Observations of direct chemical tracers of cosmic ray ionization rates, such as H$_3^+$ and He$^+$ \citep{oka06,dalgarno06,indriolo09}, could be used to confirm the cosmic ray enhancements in the immediate environment of these interacting SNRs.

\section{Summary \& Conclusions}

We have explored the class of $\gamma$-ray sources which can be explained by enhanced cosmic ray densities at the interaction sites between SNRs and molecular clouds. By correlating SNR masers with known GeV and TeV sources, we have identified an emerging class of $\gamma$-ray-bright interacting SNRs. Of the 24 known Maser SNRs currently identified there are ten with GeV to TeV-energy $\gamma$-ray associations, and six with both. If $\gamma$-ray emission from these sources is largely due to hadronic cosmic rays, the enhanced local cosmic ray ionization rates in these clouds can explain the production of OH molecules behind a C-type shock, suggested by \citet{wardle99}. Furthermore, cosmic ray ionization is typically comparable to or dominant over ionizations from interior thermal X-rays, though without detailed knowledge of the cosmic ray spectrum these results have large uncertainties. Interacting SNRs represent a promising class of $\gamma$-ray sources which are likely to be uncovered by future $\gamma$-ray observatories.


\end{document}